\documentclass[12pt,fleqn]{article}
\usepackage[dvips]{graphics}
\newcommand{\ds}{\displaystyle}

\begin{document}
\vspace*{2cm}

\noindent
{\large \bf Bound states of interacting helium atoms} \\[1em]
Stefan V. Mashkevich$^{\rm a}$
and
Stanislav I. Vilchynskyy$^{\rm b}$ \\[1em]
$^{\rm a}$Institute for Theoretical Physics,
252143 Kiev, Ukraine\footnote{Email: mash@mashke.org; sivil@ap3.bitp.kiev.ua}\\[1em]
$^{\rm b}$Chair of Quantum Field Theory, Physics Department,\\
Taras Shevchenko Kiev University, 252127 Kiev, Ukraine\footnotemark[1] \\[1em]
We study the possibility of existence of bound states for two
interacting $^4$He atoms. It is shown that for some potentials,
there exist not only discrete levels but also bands akin to those
in the Kronig--Penney model. \\[1em]
{\bf 1. INTRODUCTION} \\

It is known that most of the paradoxes of modern microscopic theory of
superfluidity
[1] have to do with the commonly accepted assumption
of the single-particle Bose condensate (SBC) prevailing in the quantum structure
of the superfluid component. In the $^4$He Bose liquid, because of
strong interaction, the SBC is strongly suppressed
(it comprises no more than 10\% of all $^4$He atoms
[2]), consequently it cannot be
the basis of the superfluid component. If in some domain of momentum space there is
strong enough attraction between atoms, bound pairs of bosons can form, whence
a pair coherent condensate appears, and it is this condensate that will be the
basis of the superfluid component.

The hypothesis of coupling of helium
atoms below the $\lambda$ point may resolve some of those paradoxes
[1,3].

The main purpose of this paper is to explore the possibility of existence of
bound states for two interacting $^4$He atoms for different theoretical and
phenomenological interatomic potentials. We will show that for some
of the potentials, discrete levels do exist. Besides, we will analyze certain
``quasi-crystallic'' models, motivated by the conjecture that at temperatures
below the $\lambda$ point, helium possesses a structure close to that of quantum
crystals. Specifically, we will consider
a model of atoms ``smeared'' over a sphere and a Kronig--Penney-type model.
These models also allow for pairing of helium atoms;
moreover, the Kronig--Penney-type model leads to a band of allowed energies.
\\[1em]
{\bf 2. TWO INTERACTING ATOMS}\\

During several decades, different phenomenological potentials of interatomic
interaction in gaseous and liquid $^4$He have been constructed, based on empirical
data on thermodynamic, kinetic and quantum mechanical properties of helium
[4--8].

Consider the Schr\"odinger equation for two helium atoms
($\Psi$ being the radial part of the wave function, $L=0$)
\begin{equation}
-\frac{\hbar^2}{2m} \frac{1}{r^2} \frac{\partial }{\partial r}
\left( r^2 \frac{\partial }{\partial r}\Psi \right)
+ \Phi(r)\Psi = E\Psi\;,
\label{Schreq}
\end{equation}
$\Phi(r)$ being the potential and $m$ the reduced mass
($2m=m_{\rm He}=6.6466\cdot10^{-24}$~g).
For a given $\Phi(r)$, we analyze Eq.~(\ref{Schreq}) for discrete levels
by means of solving it by the Runge--Kutta method from the left
and from the right and trying to choose
$\varepsilon$ such that the logarithmic derivatives of the two solutions match
at some midpoint. If this fails, it is concluded that there is
no level.\\

Table 1

Energy levels for two helium atoms assuming various interaction potentials

\begin{center}
\begin{tabular}{|l|l|l|}
\hline
Potential & $\Phi(r)$, $10^{-12}$ erg (r in \AA) & Level, K \\
\hline
\hline
Rosen--Margenau [5] &
$925 \exp(-4.4 r) - 560\exp (-5.33 r) - \frac{\ds 1.39}{\ds r^6} -
\frac{\ds 3}{\ds r^8}$
\rule[-1.1em]{0em}{2.5em}
& None \\
\hline
Slatter--Kirkwood [6] &
$770 \exp(-4.6 r) - \frac{\ds 1.49}{\ds r^6}$
\rule[-1.1em]{0em}{2.5em}
& None \\
\hline
Intem--Schneider [8] &
$1200 \exp(-4.72 r) - \frac{\ds 1.24}{\ds r^6} - \frac{\ds 1.89}{\ds r^8}$
\rule[-1.1em]{0em}{2.5em}
& None \\
\hline
Lennard--Jones &
$ 4\epsilon\left[
\left(\frac{\sigma}{r}\right)^{12}-
\left(\frac{\sigma}{r}\right)^{6}\right] $ & \\
&
Case 1 [7]:
$\sigma=2.556$ \AA, $\epsilon=10.22\;\mbox{K}$
& None \\
& Case 2 [9]:
$\sigma=2.642$ \AA, $\epsilon=10.80\;\mbox{K}$ &
0.0201 \\
\hline
Buckingham [10] &
$\left\{\begin{array}{ll}
770 \exp(-4.6 r)-\frac{\ds 1.49}{\ds r^6} \rule[-1.1em]{0em}{2.5em}
\; , \quad  r\leq 2.61;\\
977 \exp(-4.6 r)-\frac{\ds 1.50}{\ds r^6} \rule[-1.1em]{0em}{2.5em}
- \frac{\ds 2.51}{\ds r^8} \; ,
\quad r\geq 2.61
\end{array}\right. $ &
0.00632 \\
\hline
Massey--Buckingham [10] &
$1000 \exp(-4.6 r) - \frac{\ds 1.91}{\ds r^6}$
\rule[-1.1em]{0em}{2.5em} &
0.0622 \\
\hline
Buckingham--Hamilton [10] &
$977 \exp(-4.6 r) - \frac{\ds 1.5}{\ds r^6} - \frac{\ds 2.51}{\ds r^8}$
\rule[-1.1em]{0em}{2.5em} &
0.00229 \\
\hline
\end{tabular}
\end{center}
{\bf 3. AN ATOM AT THE CENTER OF A LATTICE}\\

At temperatures below the $\lambda$ point, the structure and dynamical properties
of helium are close to those of quantum crystals (e.g., Putterman
[11]
pointed out a possibility of transverse oscillations in superfluid helium).
Motivated by this, let us study the behavior of a helium atom in the
center of a lattice. For qualitative purposes, we replace the actual cubic
lattice, with the lattice constant $d=4.5$ \AA{}, by
$c=8$ atoms ``smeared'' over a sphere of radius $a=d\sqrt{3}/2$.
The potential corresponding to such a model is
\begin{equation}
U(r)=\frac{c}{2}\int_{0}^{\pi}\Phi(\sqrt{r^{2}+a^{2}-2ra\cos\theta})
\sin \theta \,d\theta \;,
\label{Umany}
\end{equation}
where $\Phi$ is taken to be the Lennard--Jones potential (case 2 above).
We then have to find a level in the potential $U(r)$, depicted below.

Obviously, since $U(r)$ is infinite at $r=a$, there will be infinitely
many levels. We find the ground level to be $E=-11.45$ K, and the
(certainly unphysical) first excited level to be $E=+35.64$ K. 

\begin{center}
\scalebox{.7}{
\includegraphics[0,0][288,180]{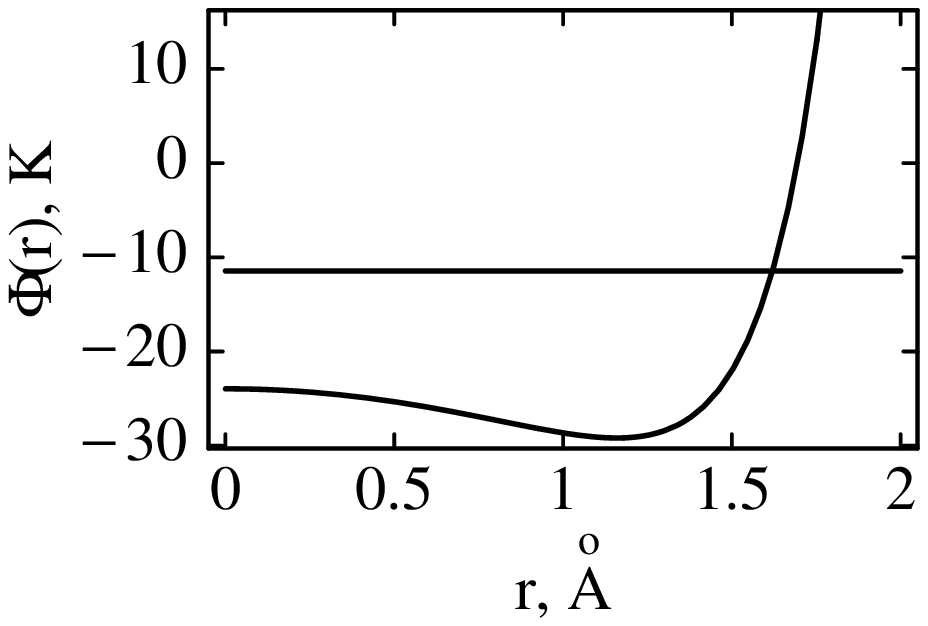}
}

Fig. 1. The ground level for the atom in the center of a lattice.
\end{center}

{\bf 4. TWO NEIGHBORING CELLS AND A BAND}\\

Now, consider a ``quasi-crystallic'' model. A
more realistic potential, without an infinite barrier, is
\begin{equation}
\Phi_2(r) = \left\{ \begin{array}{ll}
\Phi(r), & 0 < r \le a-d, \\
\Phi(a-d), & a-d < r \le a+d, \\
\Phi(2a-r), \qquad & a+d < r \le 2a.
\end{array}
\right.
\label{Phi2}
\end{equation}
The parameter $d$ is taken to be 2.2 \AA, which corresponds to twice
the quantum chemical radius of the helium atom.

We get two levels: an even one with $E=-13.862$ K and an odd one with
$E=-13.8486$ K.

\begin{center}
\scalebox{.7}{
\includegraphics[0,0][288,180]{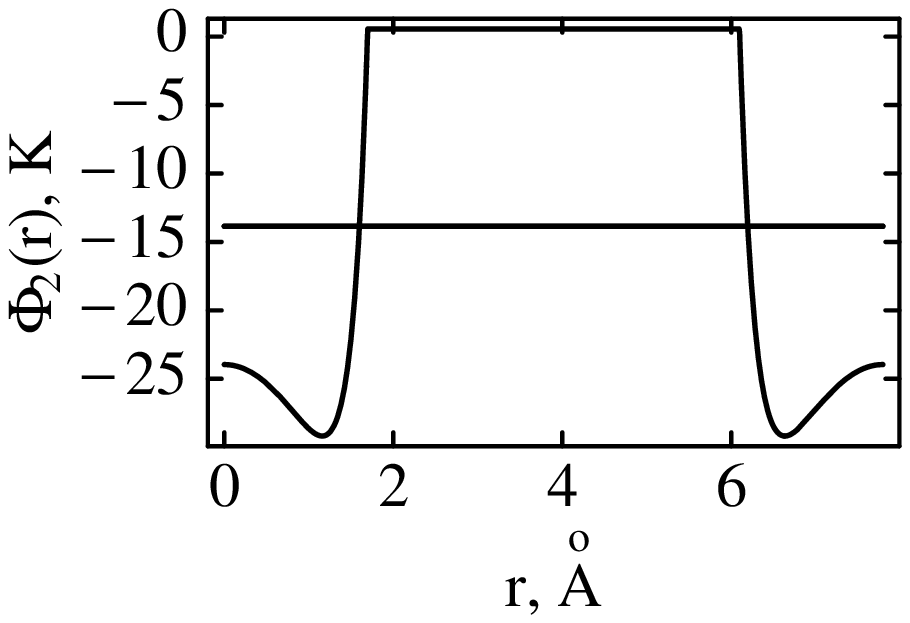}}

Fig. 2. The potential and the levels (the splitting is not noticeable) in the model of two neighboring cells.\\[1em]
\end{center}

Finally, consider a periodic potential:
\begin{eqnarray}
\Phi_P(r) & = & \Phi_2(r), \qquad 0 < r \le 2a \;; \qquad\qquad
\Phi_P(r+2a) = \Phi_P(r) \;.
\label{Phiper}
\end{eqnarray}
We find the allowed band by a well-known method 
[12]:
The wave function is represented as
\begin{equation}
\psi(r) = {\rm e}^{{\rm i}Kr}u_K(r)\;,
\label{psiper}
\end{equation}
where $u_K(r)$ is a periodic function. Matching the logarithmic derivative leads to
\begin{equation}
\cos 2Ka = \frac{[u_1(0)u'_2(a)+u_1(a)u'_2(0)]-[u_2(0)u'_1(a)+u_2(a)u'_1(0)]}
{2[u_1(0)u'_2(0)-u_2(0)u'_1(0)]}\;,
\label{cosKa}
\end{equation}
where $u_1(r)$ and $u_2(r)$ are two arbitrary linearly independent solutions
of the Schr\"odinger equation with energy $E$. The values of $E$ for which
this equation can be solved for $K$ (that is, for which the right-hand side
is not greater than 1 by absolute value), form an allowed band. Solving the
Schr\"odinger equation numerically and plotting the $E$ dependence of the
right-hand side, one finds the band and its boundaries at different values
of $d$. The maximal width of the band, at $d \simeq 2.34$, is about 1 K.
Depicted below is the potential and the band at $d=2.33$.

\begin{center}
\scalebox{0.85}{\includegraphics{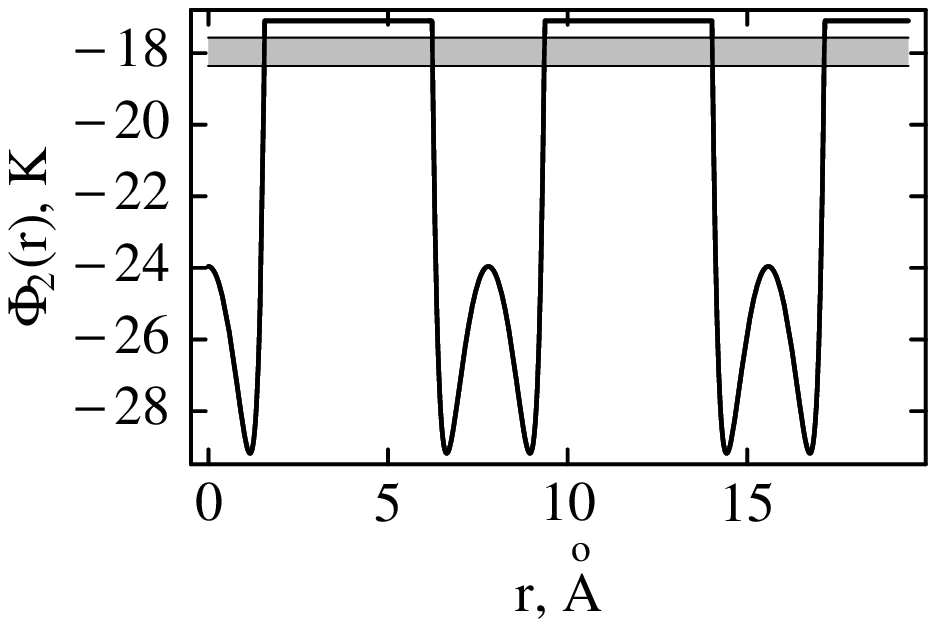}}

Fig. 3. The allowed band in the periodic potential model.
\end{center}

When $d$ is further increased, the upper edge of the band will ``overflow''
the potential and merge into the continuous spectrum. In fact, the band edges
coincide with the even and odd levels in the two-well problem considered above.
The reason is that the even and odd (with respect to $a$) wave functions of
that problem (the boundary conditions for which are $\psi(0)=\psi(2a)=0$)
can be extended into the antiperiodic ($\cos 2Ka=-1$) and periodic ($\cos 2Ka=1$)
wave functions, respectively, of the periodic potential. According to the above,
these functions correspond to the band edges.

For a rectangular potential (Kronig--Penney model
[13]) of wells of
width $a$ divided by barriers of height $V_0$ and width $b$ with the lattice
constant being fixed: $a+b=4.5$ \AA, the band is determined
by the equation
\begin{equation}
\frac{\beta^2-\alpha^2}{2\alpha\beta}\sinh \beta b \, \sin \alpha a
+ \cosh \beta b \, \cos \alpha a = \cos k(a+b)\;,
\label{Kronig}
\end{equation}
where $
\alpha = \sqrt{2mE}/\hbar,
\beta = \sqrt{2m(V_0-E)}\hbar$.
The values of $E$ for which the right-hand side is between -1 and 1,
form the allowed band.\\

Table 2

Results ($E_{\rm low}$, $E_{\rm high}$ and $\Delta E$ --- lower edge,
upper edge, and band width, respectively, all energies are in K):

\begin{center}
\begin{tabular}{|c|c|c|c|c|}
\hline
$a$ & $V_0$ & $E_{\rm low}$ & $E_{\rm high}$ & $\Delta E$ \\
\hline
2.5 & 10 & 2.15 & 2.35 & 0.20 \\
3.5 & 10 & 1.17 & 1.62 & 0.45 \\
3.5 & 5  & 0.78 & 1.56 & 0.78 \\
3.5 & 2  & 0.38 & 1.52 & 1.14 \\
4.0 & 2  & 0.20 & 1.92 & 1.72 \\
\hline
\end{tabular}
\end{center}

{\bf 5. CONCLUSION}\\

Thus, we have shown that in a superfluid Bose liquid, formation of bound pairs
of bosons is possible --- for some potentials, there are discrete levels.
Moreover, it has been shown that bound states can also exist in more
realistic models --- the one of atoms ``smeared'' over the sphere and
the periodic Kronig--Penney type model. Note in particular that the
Kronig--Penney type model allows for the existence of bands. The presence
of such bands may explain the observable critical velocities and the
energy gap for the excitation spectrum curve (``S curve'') in superfluid
helium, which will be discussed in subsequent papers.

We are grateful to P.I.~Fomin for suggesting the problem and constant
encouragement and to A.~Kostyuk for an interesting discussion and advice.
Calculations have been made using {\it Mathematica}.
\\[1em]
{\bf REFERENCES}\\[1em]
1. E.A. Pashytskiy, Fiz. Nizk. Temp., 25 (1999) 115.
\\
2. H.W.Jackson, Phys. Rev. A10 (1974) 278.
\\
3. S.I. Vilchinskii, P.I. Fomin, Sov. J. Low Temp. Phys., 21 (1995) 7.
\\
4. E.A. Mason, Journ. Chem. Phys., 22 (1954) 1678.
\\
5. P. Rosen, Journ. Chem. Phys., 18 (1950) 1182;
H. Margenau, Phys. Rev., 56 (1939) 1000;
C.H. Page, Phys. Rev., 53 (1938) 426.
\\
6. J.C. Slatter, J.G. Kirkwood, Phys. Rev., 37 (1931) 682.
\\
7. J.O. Hirschfelder, Ch.F. Curtiss, R.B. Bird,
Molecular Theory of Gases and Liquids, New York, 1954.
\\
8. I.G. Kaplan, Introduction to the Theory of
Intermolecular Interaction, Nauka, Moscow, 1982.
\\
9. B. Plushkin (ed.), Molecular interaction from the diatomic molecules to the polymer, Mir, Moscow, 1981.
\\
10. R.A. Buckingham, J. Hamilton, H.S.W. Massey, Proc. Roy. Soc.,
A179 (1941) 103.
\\
11. S. Putterman, Superfluid Hydrodynamics,
American Elsevier Publishing Company, Inc, New York, 1974.
\\
12. S. Fluegge, Problems on Quantum Mechanics, Mir, Moscow, 1974,
vol. 1., p. 75.
\\
13. R. Kronig, W. Penney, Proc. Roy. Soc., 130 (1931) 499.
\end{document}